\documentclass[12pt]{article}

\ifx\pdfoutput\undefined
\usepackage[dvips,bookmarks]{hyperref}
\else
\usepackage{hyperref}
\fi
\hypersetup{colorlinks=false,bookmarksopen,bookmarksnumbered,citecolor=blue,
   pdfstartview=FitH}

\usepackage{latexsym}
\usepackage{marginnote}
\usepackage{amssymb,amsfonts,amsmath}
\usepackage{graphicx}
\usepackage{indentfirst}
\usepackage{bbm}
\usepackage{colortbl}
\usepackage{hhline}

\parskip=\medskipamount

\def\a{\alpha}

\def\b{\beta}

\def\d{\delta}
\def\e{\epsilon}

\def\g{\gamma}
\def\G{\Gamma}

\def\l{\lambda}
\def\m{\mu}
\def\n{\nu}

\def\p{\pi}
\def\q{\theta}
\def\r{\rho}

\newcommand{\ve}{\varepsilon}                            %new
\newcommand{\be}{\begin{equation}}
\newcommand{\ee}{\end{equation}}
\newcommand{\bea}{\begin{eqnarray}}
\newcommand{\eea}{\end{eqnarray}}

\def\double #1{#1{\hbox{\kern-2pt $#1$}}}

\newcommand{\W}{\Omega}
\newcommand{\no}{\nonumber}
\newcommand{\ba}{\bar}
\newcommand{\ii}{\mathrm{i}}

\newcommand{\gc}[2]{\left. \lbrack #1, #2\} \right.}

\newcommand{\bsubeq}{\begin{subequations}}
\newcommand{\esubeq}{\end{subequations}}

\begin{document}

\begin{titlepage}

\vspace{2cm}

\begin{center}
{\Large \bf Deriving all $p \,$-brane superalgebras via integrability }
\end{center}

\vspace{2cm}

\begin{center}

{\large

D. T. Grasso\footnote{{darren.grasso@uwa.edu.au}},
I. N. McArthur\footnote{{ian.mcarthur@uwa.edu.au}}
} \\
\vspace{5mm}

\footnotesize{
{\it School of Physics M013,
The University of Western Australia\\
35 Stirling Highway, Crawley W.A. 6009 Australia }}
~\\

\end{center}
\vspace{3cm}

\begin{abstract}
\baselineskip=14pt
\noindent
\noindent
In previous work we demonstrated that the enlarged super-Poincare algebras which underlie $p$-brane and $D$-brane actions in superstring theory can be directly determined based on the integrability of supersymmetry transformations assigned to fields appearing in Wess-Zumino terms.  In that work we derived $p$-brane superalgebras for $p=2$ and 3.  Here we extend our previous results and give a compact expression for superalgebras for all valid $p$.
\end{abstract}
\vspace{1cm}

\vfill
\end{titlepage}

\tableofcontents{}
\vspace{1cm}
\bigskip\hrule

\section{Introduction and Background}
\setcounter{equation}{0}

The actions for $p$-branes generalize the Green-Schwarz action \cite{Green:1983sg} for superstrings from a two dimensional worldvolume embedded into a background superspace to $(p+1)$-dimensional worldvolumes. In particular, they contain a Wess-Zumino term generalizing that for the superstring \cite{Henneaux:1984mh},  whose presence is required to realize a local fermionic symmetry (kappa symmetry) necessary to ensure worldvolume supersymmetry \cite{de Azcarraga:1982dw, de Azcarraga:1982xk, Siegel:1983hh}.  Suitable Wess-Zumino terms exist only when $p$-branes are embedded in  superspaces with spacetime dimensions $D$ for which the  supersymmetry invariant $(p+2)$-form\footnote{We suppress all $\wedge$ symbols when multiplying forms.}
\begin{equation}
  h^{(p+2)}=\p^{a_1}\cdots\p^{a_p}(d\ba{\q}\G_{a_1}\cdots\G_{a_p}d\q) \label{hdef}
\end{equation}
is closed \cite{Achucarro:1987nc, Evans:1988jb}. Our conventions are that the $D$-dimensional $\cal{N}$ = 1 supersymmetry algebra is
\begin{equation}
  \left\{Q_\a,Q_\b\right\}=-2(C\G^a)_{\a\b}P_a\,,
\end{equation}
and superspace coordinates transform as
\begin{align}
  \d_\e x^a&=\ii (\ba{\e}\G^a\q) \\
  \d_\e \q^\a&=\e^\a
\end{align}
under supersymmetry transformations.  The one-forms
\begin{equation}
  \p^a=d x^a-\ii (\ba{\q}\G^a d \q), \, \, d \q^\a
\end{equation}
are invariant under supersymmetry transformations.

The Wess-Zumino term in the $p$-brane action  is constructed as the pull-back to the worldvolume of a superspace form $b^{(p+1)}$ determined by
\begin{equation}\label{bdef}
  h^{(p+2)}= d b^{(p+1)},
 \end{equation}
 consistent with the closure of $ h^{(p+2)}$ and the triviality of the cohomology of flat superspace.
 The form $ h^{(p+2)}$ is non-vanishing  only for values of $p$ and spacetime dimension $D$ for which $(C \G_{a_1\cdots a_p})_{\a\b}$ is symmetric in its spinor indices, where $\G_{a_1\cdots a_p}$ is an anti-symmetrized product of $D$-dimensional gamma matrices (note that this vanishes for $p > D$).
 Closure of $ h^{(p+2)}$ requires the gamma matrix identities\footnote{Due to the symmetry of $(C\G^{a})_{\a\b}$, for $p=1$ this expression is equivalent to that which usually appears in the literature where only thee of the four spinor indices are symmetrized.}
\begin{equation}
   0=(C\G^{a_1})_{(\a\b}(C \G_{a_1\cdots a_p})_{\g\d)}\,. \label{gammaid1}
\end{equation}
Here, the round brackets on spinor indices denote symmetrisation\footnote{Note that our (anti)symmetrization of $n$ indices carries a factor of $1/n!$}.  The existence of these identities places further restrictions on allowed values of $p$ for a given spacetime dimension $D$, and results in the ``$p$-brane scan''  \cite{Achucarro:1987nc}, which is reproduced in Table \ref{table} below. It turns out that the closure of $ h^{(p+2)}$ ensures equal numbers of bosonic and fermionic degrees of freedom on the worldvolume \cite{Achucarro:1987nc}, a necessary requirement for worldvolume supersymmetry.

\begin{table}[h!]
  \begin{center}
\scalebox{0.8}{\begin{tabular}{cc|c|c|c|c|c|c|c|c|c}
  \hline
  $\quad D\quad$ & \multicolumn{10}{l}{$p$} \\\cline{2-11}
   & $\quad 1\quad$ & $\quad 2\quad$ & $\quad 3\quad$ &$\quad 4\quad$ & $\quad 5\quad$ & $\quad 6\quad$ & $\quad 7\quad$ & $\quad 8\quad$ & $\quad 9\quad$ & $\quad 10\quad$\\\hline
  3 & $\checkmark$ &   &   &   &   &   &   &   &   \\\cline{2-11}
  4 & $\checkmark$ & $\checkmark$ &    &   &   &   &   &   &   &   \\\cline{2-11}
  5 &   & $\checkmark$ &   &   &   &   &   &   &   &   \\\cline{2-11}
  6 & $\checkmark$ &   & $\checkmark$ &   &   &   &   &   &   &   \\\cline{2-11}
  7 &   & $\checkmark$ &   &   &   &   &   &   &   &   \\\cline{2-11}
  8 &   &   & $\checkmark$ &   &   &   &   &   &   &   \\\cline{2-11}
  9 &   &   &   & $\checkmark$ &   &   &   &   &   &   \\\cline{2-11}
  10 & $\checkmark$ &   &   &   & $\checkmark$ &   &   &   &   &   \\\cline{2-11}
  11 &   & $\checkmark$ &   &   &   &   &   &   &   &   \\\cline{2-11}
  12 &   &   &   &   &   &   &   &   &   &   \\\hline
\end{tabular}}
  \caption{The ``$p$-brane scan'' from \cite{Achucarro:1987nc}. The tick marks indicate the allowed values of $p$ and $D$ in this paper.}
    \label{table}
\end{center}
\end{table}

The superspace form $b^{(p+1)}$ cannot be chosen to be invariant under supersymmetry transformations \cite{DeAzcarraga:1989vh}, as a result of which the Wess-Zumino Lagrangian is quasi-invariant. This   leads to a modification of the algebra of Noether charges resulting in an enlarged supersymmetry algebra \cite{de Azcarraga:1989gm, DeAzcarraga:1991tm}. Enlarged superalgebras were also encountered in approaches that seek to construct a manifestly supersymmetric Wess-Zumino term \cite{Siegel:1994xr} by postulating a  $p$-form $A^{(p)}$ whose supersymmetry transformations are determined  by the requirement that
\begin{align}
\mathcal{F}^{(p+1)}=d A^{(p)}-b^{(p+1)}
\end{align}
is invariant. The construction of $A^{(p)}$ requires the introduction of an enlarged superspace related by the coset space construction to enlarged superalgebras similar to those appearing in the enlarged algebra of Noether charges \cite{Siegel:1994xr, Bergshoeff:1989ax, Bergshoeff:1995hm, Sezgin:1996cj}. Stated technically, $h^{(p+2)}$ belongs to a trivial Chevalley-Eilenberg cohomology class for the enlarged superspace.

In \cite{McArthur:2015pna} and \cite{grassomcarthur}, we used an alternative approach to systematically determine the expanded superalgebras associated with $p$-branes, namely by requiring integrability of the supersymmetry transformations assigned to $A^{(p)}$ in solving the  cohomology problem
\be
h^{(p+2)} = d b^{(p+1)}, \quad \d_{\e} b^{(p+1)} = d \d_{\e} A^{(p)}.
\ee
We were able to systematically derive the enlarged superalgebras that underly $p$-branes for the cases $p=1$, $p=2,$ $p=3,$ as well as the $D$-brane worldvolume one-form and the $M5$ brane two-form. In this paper, we extend these results to all possible combinations of $p$ and $D$ for which the form $h^{(p+2)}$ exists and is closed via the  gamma matrix identity (\ref{gammaid1}), namely for the  ``$p$-brane scan'' in Table \ref{table}.

The advantage of our approach is that it is systematic and allows derivation of the most general superalgebras for all allowed values of $p$ and $D$. Previously, the most comprehensive approach has been via free differential superalgebras \cite{Bergshoeff:1995hm, Chryssomalakos:1999xd}. There, the numerical coefficients that appear in front of the terms on the right hand side of the (anti)commutators are listed on a case-by-case basis with no general structure; here, we are able to give an expression for these coefficients in closed form valid for all allowed values of $p$. Quite surprisingly, they turn out to involve the Euler gamma function for non-integer values of its argument.

The overall calculation presented here is notationally complicated and quite long. For clarity we now summarise the major steps involved, around which we organise the layout of this paper.

In Section \ref{sec:notation} we outline the new notational devices that are required for handling arbitrary $p$ calculations. Section \ref{sec:b} details how we solve
\begin{align}
  h^{(p+2)}=d b^{(p+1)}
\end{align}
to find an expression for $b^{(p+1)}$ which involves a number of arbitrary `constants of integration'.  In Section \ref{sec:dA} we use our expression for $b^{(p+1)}$ and solve
\begin{equation}
 \d_{\e} b^{(p+1)} = d \d_{\e} A^{(p)}\,
\end{equation}
to find an expression for $\d_{\e} A^{(p)},$ which involves the introduction of more constants of integration.

In Section \ref{sec:comvarA} we compute $[\d_{\e_2},\d_{\e_1}] A^{(p)}$, the commutator of two supersymmetry transformations of $A^{(p)}$.  We then use the result, along with the freedom provided by the constants of integration, to generate an expression which is consistent with a (non-central) extension of the supersymmetry algebra.

Section \ref{sec:computingthealgebra} details how we use the results of Section \ref{sec:comvarA} to determine the entire algebra.  In Section \ref{sec:thealgebra} we collect together and display our final result: the extended supersymmetric algebra for any  $p$ in which the gamma matrix identity (\ref{gammaid1}) applies.

Using a coset construction, in Section \ref{sec:F} we use all of our previous results to determine a manifestly supersymmetric expression for $\mathcal{F}^{(p+1)}=d A^{(p)}-b^{(p+1)}$ in terms of Maurer-Cartan forms on an extended superspace.

Finally, Section \ref{sec:conclusion} contains some concluding comments.

\section{Notation}\label{sec:notation}
\setcounter{equation}{0}

To be able to handle the case of arbitrary $p$ it will be convenient to introduce a number of notational devices.

Firstly, as seen in the definition of $h^{(p+2)}$, we will be required to manage products involving an arbitrary number of forms.  To this end we introduce the notation:
\begin{align}
  K^a:=-\ii (\ba{\q}\G^a d \q)\,,\qquad K_{a_1\cdots a_p}:=(\ba{\q}\G_{a_1\cdots a_p} d \q)\,,
\end{align}
and the following shorthand for an ordered product of $n-m+1$ factors of any object $L^A$ carrying some index $A$ (Lorentz or spinor)
\begin{equation}
  (L^A)^{m\rightarrow n}:= L^{A_m}L^{A_{m+1}}L^{A_{m+2}} \cdots L^{A_{n-1}} L^{A_n}
\end{equation}
where $n > m$ are both positive integers labeling the index of each factor.  Furthermore, in the case where $n=m$
we define
 \begin{equation}
  (L^A)^{m\rightarrow m}:= L^{A_m}
\end{equation}
and in the case where $n-m=-1$ we define
\begin{equation}
  (L^A)^{m\rightarrow m-1}:= 1\,.
\end{equation}
For clarity, some explicit examples of the usage of this device include:
\begin{align}
  (dx^a)^{1\rightarrow 4}= dx^{a_1}dx^{a_2}dx^{a_3} dx^{a_4} \,,\quad (d\q^\a)^{2\rightarrow 4}= d\q^{\a_2}d\q^{\a_3}d\q^{\a_4}\,, \quad (d x^a)^{3\rightarrow 3}=d x^{a_3} \\   (d x^a)^{6\rightarrow 5}= 1\,,\quad (K^b)^{3\rightarrow 5}= K^{b_3}K^{b_4}K^{b_5}= (-\ii)^3 (\ba{\q}\G^{b_3} d \q)(\ba{\q}\G^{b_4} d \q)(\ba{\q}\G^{b_5} d \q)\,.
\end{align}
Often we will find the need to (anti)symmetrize indices appearing in this new notation.  This will be achieved by invoking square or round brackets in the obvious way, as for example:
\begin{align}
  (dx^{[a})^{1\rightarrow 3}(K^{a]})^{4\rightarrow 5}&=dx^{[a_1}dx^{a_2}dx^{a_3}K^{a_4}K^{a_5]}\,, \quad
  \q^{(\a_1}(d\q^{\a)})^{2\rightarrow 4}= \q^{(\a_1}d\q^{\a_2}d\q^{\a_3}d\q^{\a_4)}.
\end{align}

We also introduce the notation $a_{m\rightarrow n}$ as shorthand for a string of $n-m+1$ indexed indices, used, for example, as follows:
\begin{equation}
 K_{a_{1\rightarrow p}}=K_{a_1 a_2\cdots a_{p-1} a_p}\,, \qquad K_{b c a_{m\rightarrow n}}=K_{b c a_m a_{m+1}\cdots a_{n-1} a_n}
\end{equation}
with $n>m$ positive integers.  We also extend this notation to include the case where $n=m$ which indicates that there is just a single indexed index as follows
\begin{equation}
 K_{b c a_{m\rightarrow m}}=K_{b c a_m}\,.
\end{equation}
The case where $n-m=-1$ will indicate the absence of  an index, for example
\begin{equation}
 K_{b c a_{m\rightarrow m-1}d}=K_{b c d}\,.
\end{equation}
The need for and usage of these notational devices will become apparent as we proceed.

For later purposes we will also define
\begin{align}
  M_\e^a:=\ii (\ba{\e}\G^a \q),
\end{align}
so that under a supersymmetry transformation with parameter $\e^\a$ we have
\begin{align}
  \d_\e x^a =M_\e^a\,, \qquad \d_\e K^a =-d M_\e^a\,.
\end{align}

\section{Computing $b^{(p+1)}$}\label{sec:b}
\setcounter{equation}{0}

Using the notation introduced above we can express the $p+2$ form $h^{(p+2)}$ as
 \begin{align}
  h^{(p+2)}&=\p^{a_1}\cdots\p^{a_p}(d\ba{\q}\G_{a_1}\cdots\G_{a_p}d\q)\no\\
  &=(\p^a)^{1\rightarrow p}d K_{a_{1\rightarrow p}} \no\\
  &=(d x^a+K^a)^{1\rightarrow p}d K_{a_{1\rightarrow p}}\no\\
  &=\sum_{n=0}^p\frac{p!}{n!(p-n)!}(dx^a)^{1\rightarrow n}(K^a)^{n+1\rightarrow p}dK_{a_{1\rightarrow p}}\label{h} \, ,
\end{align}
having used the binomial theorem after noting that $K^b dx^a =-dx^a K^b$, the indices of which are contracted with the anti-symmetrized product of gamma matrices.

To find an expression for $b^{(p+1)}$ we are required to solve \eqref{bdef}.  We achieve this by first writing down the most general possible expression for $b^{(p+1)}$.  This will be an arbitrary linear combination -- modulo any equivalent terms due to the gamma matrix identities \eqref{gammaid1} -- of all non-trivial $p+1$ forms constructed from factors of $x^a$, $d x^a$, $K^a$ and $dK^a$, all of which are either contracted with a single factor of $K_{a_{1\rightarrow p}}$ or a single factor of $dK_{a_{1\rightarrow p}}$. Noting that \eqref{gammaid1} implies
\begin{align}
  K^{a_1} dK_{a_{1\rightarrow p}}=-d K^{a_1} K_{a_{1\rightarrow p}}\,, \label{sgi}
\end{align}
then this most general expression for $b^{(p+1)}$ is
\begin{align}
   b^{(p+1)}=\,&\,\sum_{n=0}^p b_n (dx^a)^{1\rightarrow n}(K^a)^{n+1\rightarrow p}K_{a_{1\rightarrow p}}+\sum_{n=1}^p c_n x^{a_1}(dx^a)^{2\rightarrow n}(K^a)^{n+1\rightarrow p}d K_{a_{1\rightarrow p}}
\end{align}
where $b_0,b_1,\ldots b_p,c_1,c_2,\ldots ,c_p$ is a collection of unknown constants.   From this expression it follows that
\begin{multline}
  d b^{(p+1)}=\sum_{n=0}^p (-1)^n (-1)^{p-n}(p-n+1) b_n (dx^a)^{1\rightarrow n}(K^a)^{n+1\rightarrow p}dK_{a_{1\rightarrow p}}
  \\+\sum_{n=1}^p c_n d x^{a_1}(dx^a)^{2\rightarrow n}(K^a)^{n+1\rightarrow p}dK_{a_{1\rightarrow p}} \label{db}
\end{multline}
having exploited the identity \eqref{gammaid1} a number of times in the form of \eqref{sgi} and
\begin{align}
  d K^{a_1} dK_{a_{1\rightarrow p}}=0\,. \label{sgi1}
\end{align}
To facilitate a comparison with \eqref{h}, we can write our result \eqref{db} as
\begin{align}
  d b^{(p+1)}=\,&\,\sum_{n=0}^p \Big(c_n+(-1)^p(p-n+1) b_n\Big) (dx^a)^{1\rightarrow n}(K^a)^{n+1\rightarrow p}dK_{a_{1\rightarrow p}}\,,\label{computedb}
\end{align}
having collected all terms together under a single summation via the introduction of an additional constant $c_0$, which is in fact just zero.

A comparison of \eqref{computedb} and \eqref{h} now immediately shows that if our general expression for $b^{(p+1)}$ is to provide us with a solution to \eqref{bdef}, we require the coefficients $b_n$ and $c_n$ satisfy
\begin{align}
  c_n+(-1)^p(p-n+1) b_n=\frac{p!}{n!(p-n)!}\qquad\qquad n=0,1,2,\ldots ,p\,. \label{cb}
\end{align}
Since $c_0=0$ we find that
\begin{align}
  b_0=\frac{(-1)^p}{p+1}\,.
\end{align}

Using the relations \eqref{cb} to eliminate the constants $c_n$, we conclude that
\begin{align}
   b^{(p+1)}=\,&\,\sum_{n=0}^p b_n (dx^a)^{1\rightarrow n}(K^a)^{n+1\rightarrow p}K_{a_{1\rightarrow p}}+\sum_{n=1}^p c_n x^{a_1}(dx^a)^{2\rightarrow n}(K^a)^{n+1\rightarrow p}d K_{a_{1\rightarrow p}} \label{bresult}
\end{align}
with
\begin{align}
  c_0=0\,, \quad c_n=\frac{p!}{n!(p-n)!}-(-1)^p(p-n+1) b_n\,  \qquad n=0,1,\ldots ,p\,. \label{c}
\end{align}
Thus, in solving for $b^{(p+1)}$, we find that the most general solution contains $p$ arbitrary `constants of integration' $b_1,b_2,\ldots b_p$. Note that a general expression for  $b^{(p+1)}$ was provided in \cite{Evans:1988jb}; however, the author required that the expression be spacetime translation invariant (i.e. no explicit $x^a$ dependence). Taking into account notational differences, our result can be made to coincide with that in \cite{Evans:1988jb} by an appropriate choice of integration constants. As  noted in \cite{Hammer:1997ts, Reimers:2005jf}, restriction to spacetime translation invariant expressions results in enlarged superalgebras which are not the most general possible. In particular, some fermionic generators are missed.

\section{Computing $\d_\e A^{(p)}$} \label{sec:dA}
\setcounter{equation}{0}

We now wish to find an expression for $\d_{\e} A^{(p)}$ using
\begin{equation}
 \d_{\e} b^{(p+1)} = d \d_{\e} A^{(p)}\,, \label{ap}
\end{equation}
where $\d_\e$ denotes a supersymmetry transformation with parameter $\e^\a$.

With a little work, including a change of variables in some of the summations, it can be shown that the supersymmetry transformation of expression \eqref{bresult} is given by\footnote{Throughout this paper it is to be understood that if the lower limit of a summation exceeds the upper limit then that summation formally vanishes, making no contribution to the expression.}
\begin{align}
  \d_\e &b^{(p+1)} =\sum_{n=1}^p \Big(n b_n+(n-p-1)b_{n-1}\Big) d M_{\e}^{a_1} (dx^a)^{2\rightarrow n}(K^a)^{n+1\rightarrow p}K_{a_{1\rightarrow p}}\no\\&+\sum_{n=2}^p \Big((n-1)c_n+(n-p-1)c_{n-1}\Big) x^{a_1}d M_{\e}^{a_2} (dx^a)^{3\rightarrow n}(K^a)^{n+1\rightarrow p}d K_{a_{1\rightarrow p}}\no\\&+\sum_{n=0}^p b_n (dx^a)^{1\rightarrow n}(K^a)^{n+1\rightarrow p}\d_\e K_{a_{1\rightarrow p}}+
  \sum_{n=1}^p c_n M_{\e}^{a_1} (dx^a)^{2\rightarrow n}(K^a)^{n+1\rightarrow p}dK_{a_{1\rightarrow p}}\, \label{varb}
\end{align}
with $M_\e^a=\ii (\ba{\e}\G^a \q)$.

To find the $p$-form $\d_{\e} A^{(p)}$ appearing in \eqref{ap}, we follow what is essentially the same procedure that provided us with $b^{(p+1)}$ earlier. We begin by writing write down the most general expression for $\d_{\e} A^{(p)}$ modulo terms which are equivalent due to the gamma matrix identities. This general expression is just an arbitrary linear combination of all possible non-trivial $p$-forms, linear in $\e$, and constructed from either:
\begin{enumerate}
  \item factors of $x^a$, $d x^a$, $K^a$ and $dK^a$, all contracted with a single factor of $(\ba{\e}\G_{a_{1\rightarrow p}}\q)$;
  \item factors of $x^a$, $d x^a$, $K^a$ and $dK^a$, all contracted with a single factor of $(\ba{\e}\G_{a_{1\rightarrow p}}d\q)$;
  \item factors of $x^a$, $d x^a$, $K^a$ $dK^a$ and $(\ba{\e}\G^{a}\q)$, all contracted with a single factor of $K_{a_{1\rightarrow p}}$ or $dK_{a_{1\rightarrow p}}$;
  \item factors of $x^a$, $d x^a$, $K^a$, $dK^a$ and $(\ba{\e}\G^{a}d\q)$, all contracted with a single factor of $K_{a_{1\rightarrow p}}$ or $dK_{a_{1\rightarrow p}}$.
\end{enumerate}
Since the gamma matrix identities imply \eqref{sgi} and
\begin{align}
  (d\ba{\q}\G^{a_p} d\q)(\ba{\e}\G_{a_{1\rightarrow p}}\q)+(\ba{\e}\G^{a_p} \q)(d\ba{\q}\G_{a_{1\rightarrow p}}d\q)= 2(\ba{\e}\G^{a_p} d\q)(\ba{\q}\G_{a_{1\rightarrow p}}d\q)-2(\ba{\q}\G^{a_p} d\q)(\ba{\e}\G_{a_{1\rightarrow p}}d\q)\,, \label{sgi2}
\end{align}
it follows that the most general expression for $\d_{\e} A^{(p)}$ is given by
\begin{align}
  \d_{\e} A^{(p)}=&\sum_{n=0}^p e_n (dx^a)^{1\rightarrow n}(K^a)^{n+1\rightarrow p}(\ba{\e}\G_{a_{1\rightarrow p}}\q) \no
  \\+&\sum_{n=1}^p f_n x^{a_1}(dx^a)^{2\rightarrow n}(K^a)^{n+1\rightarrow p} (\ba{\e}\G_{a_{1\rightarrow p}}d\q) \no
  \\+&\sum_{n=1}^{p-1} g_n x^{a_1}(dx^a)^{2\rightarrow n}(\ba{\e}\G^{a_{n+1}}d\q)(K^a)^{n+2\rightarrow p} K_{a_{1\rightarrow p}}  \no
  \\+&\sum_{n=1}^{p-1} h_n x^{a_1}(dx^a)^{2\rightarrow n}(\ba{\e}\G^{a_{n+1}}\q)(K^a)^{n+2\rightarrow p} dK_{a_{1\rightarrow p}} \no
  \\+&\sum_{n=0}^{p-1} k_n (dx^a)^{1\rightarrow n}(\ba{\e}\G^{a_{n+1}}\q)(K^a)^{n+2\rightarrow p} K_{a_{1\rightarrow p}} \label{defae}
\end{align}
where we have introduced the collection of unknown constants $\{e_n\}_{n=0}^p$, $\{f_n\}_{n=1}^p$, $\{g_n,h_n\}_{n=1}^{p-1}$ and $\{k_n\}_{n=0}^{p-1}$. Taking the exterior derivative of this yields
\begin{align}
   d \d_{\e} A^{(p)}\,&\,=\sum_{n=1}^p\Big(2(-1)^p(p-n+1) e_{n-1}-\ii g_{n-1}+\ii (-1)^n k_{n-1}\Big)  d M_{\e}^{a_1} (dx^a)^{2\rightarrow n}(K^a)^{n+1\rightarrow p}K_{a_{1\rightarrow p}}   \no \\
   &+\sum_{n=2}^p\Big((-1)^p (p-n+1) f_{n-1}+\ii(-1)^p(p-n+1) g_{n-1}+\ii (-1)^{n-1}h_{n-1}\Big)\no \\ &\qquad  \qquad\qquad\qquad\qquad \qquad\qquad \qquad\qquad \qquad \times x^{a_1} d M_{\e}^{a_2} (dx^a)^{3\rightarrow n}(K^a)^{n+1\rightarrow p}dK_{a_{1\rightarrow p}}\no
   \\&+\sum_{n=0}^p\Big((-1)^p \big(2 p-2 n+1\big)e_{n}+ f_{n}\Big)  (dx^a)^{1\rightarrow n}(K^a)^{n+1\rightarrow p}\d_\e K_{a_{1\rightarrow p}}
    \no \\&+\sum_{n=1}^p\Big((p-n+1)e_{n-1}+\ii (-1)^n h_{n-1}+\ii(-1)^{p+n+1}(p-n+1) k_{n-1}\Big)\no \\ &\qquad  \qquad\qquad\qquad\qquad \qquad\qquad \qquad\qquad \qquad \times M_\e^{a_1} (dx^a)^{2\rightarrow n}(K^a)^{n+1\rightarrow p}d K_{a_{1\rightarrow p}} \label{da}
\end{align}
where, to facilitate a direct comparison with \eqref{varb}, we have introduced the additional constants $f_0=g_0=h_0=0$.  In deriving the above expression for $d \d_{\e} A^{(p)}$ the gamma matrix identity \eqref{gammaid1} has been used in the form of \eqref{sgi2}
and
\begin{align}
  (d\ba{\q}\G^{a_p} d\q)(\ba{\e}\G_{a_{1\rightarrow p}}d\q)=-(\ba{\e}\G^{a_p} d\q)(d\ba{\q}\G_{a_{1\rightarrow p}}d\q)\,.
\end{align}
A comparison of  \eqref{da} and \eqref{varb} now immediately implies the following relationships between the constants $b_n$, $e_n$, $f_n$, $g_n$, $h_n$ and $k_n$:
\begin{align}
  n b_n+(n-p-1)b_{n-1}&=2(-1)^p(p-n+1) e_{n-1}\no \\&\hspace{3cm}-\ii g_{n-1}+\ii (-1)^n k_{n-1}\, \quad n=1,2,\ldots,p
\end{align}
\begin{align}
 (n-1)c_n+&(n-p-1)c_{n-1} =(-1)^p (p-n+1) f_{n-1}\no \\&\hspace{1cm}+\ii(-1)^p(p-n+1) g_{n-1}+\ii (-1)^{n-1}h_{n-1}\, \quad n=2,3,\ldots,p
\end{align}
\begin{align}
 b_n = (-1)^p \big(2p-2n+1\big)e_{n}+ f_{n}\, \quad n=0,1,2,\ldots,p
\end{align}
and
\begin{align}
 c_n &= (p-n+1)e_{n-1}+\ii (-1)^n h_{n-1}\no \\ &\hspace{3cm}+\ii(-1)^{p+n+1}(p-n+1) k_{n-1}\,  \quad n=1,2,3,\ldots,p
\end{align}
Along with \eqref{c}, a number of these relationships can be used to eliminate some of the constants in favour of others, while the remainder form consistency conditions. We ultimately conclude that
\begin{align}
  \d_\e A^{(p)}=\,&\,\sum_{n=0}^p e_n (dx^a)^{1\rightarrow n}(K^a)^{n+1\rightarrow p}(\ba{\e}\G_{a_{1\rightarrow p}}\q) \no
  \\&+\sum_{n=1}^p f_n x^{a_1}(dx^a)^{2\rightarrow n}(K^a)^{n+1\rightarrow p} (\ba{\e}\G_{a_{1\rightarrow p}}d\q) \no
  \\&+\sum_{n=1}^{p-1} g_n x^{a_1}(dx^a)^{2\rightarrow n}(\ba{\e}\G^{a_{n+1}}d\q)(K^a)^{n+2\rightarrow p} K_{a_{1\rightarrow p}}  \no
  \\&+\sum_{n=1}^{p-1} h_n x^{a_1}(dx^a)^{2\rightarrow n}(\ba{\e}\G^{a_{n+1}}\q)(K^a)^{n+2\rightarrow p} dK_{a_{1\rightarrow p}} \no
  \\&+\sum_{n=0}^{p-1} k_n (dx^a)^{1\rightarrow n}(\ba{\e}\G^{a_{n+1}}\q)(K^a)^{n+2\rightarrow p} K_{a_{1\rightarrow p}} \label{varAe}
\end{align}
with
\begin{align}
  e_0&=\frac{1}{(1+p)(1+2p)}\,, \label{e}\\ \no \\
  f_n&=b_n + (-1)^{p+1} (2 p - 2 n + 1) e_n\, \qquad\qquad n=1,2,\ldots, p  \label{f} \\ \no \\
  g_n&=\ii (n - p) b_n + \ii n b_{n+1} -\ii  (-1)^p (2 p - 2 n + 1) e_n \no \\
  &\hspace{4cm}+ \frac{(-1)^{n+p}}{n - p} h_n+ \frac{\ii (-1)^p p!}{(p-n)!(n+1)!}\, \qquad  n=1,2,\ldots, p-1 \label{g}\\ \no \\
  k_n&=\ii (-1)^n b_{n+1} +\ii (-1)^{p-n} e_n +  \frac{(-1)^{p}}{p - n}h_n - \frac{\ii (-1)^{p+n}  p!}{(p-n)!(n+1)!}\, \quad n=0,1,\ldots, p-1\,. \label{k}
\end{align}
And so we have now derived an expression for $\d_\e A^{(p)}$ in terms of $3p-1$ arbitrary constants of integration: $b_1, b_2, \ldots, b_p, h_1, h_2, \ldots, h_{p-1}, e_1, e_2, \ldots e_p\,$.

\section{Computing $[\d_{\e_2},\d_{\e_1}] A^{(p)}$}\label{sec:comvarA}
\setcounter{equation}{0}

We now wish to determine $[\d_{\e_2},\d_{\e_1}] A^{(p)}$, the commutator of two supersymmetry transformation of $A^{(p)}$.  Proceeding directly by taking a second variation of the expression \eqref{varAe} -- where the constants are subject to \eqref{e}-\eqref{k} -- we obtain the following result:
\begin{align}
  [\d_{\e_2}&,\d_{\e_1}] A^{(p)}=2\sum_{n=1}^{p-1}h_n x^{a_1}(dx^a)^{2\rightarrow n}(\ba{\e}_1\G^{a_{n+1}}\e_2)(K^a)^{n+2\rightarrow p}d K_{a_{1\rightarrow p}}  \marginnote{(1)}  % line 9 from notes page 30
  \no \\
  &+2\sum_{n=0}^{p-1}k_n (dx^a)^{1\rightarrow n}(\ba{\e}_1\G^{a_{n+1}}\e_2)(K^a)^{n+2\rightarrow p} K_{a_{1\rightarrow p}}  \marginnote{(2)} % line 10 from notes
  \no \\
  &+2\sum_{n=0}^pe_n (dx^a)^{1\rightarrow n}(K^a)^{n+1\rightarrow p}(\bar{\e}_{1}\G_{a_{1\rightarrow p}}\e_2) \marginnote{(3)} % line 2 from notes
   \no \\
&+2\sum_{n=2}^p\Big((n-1)f_n+(n-p-1)f_{n-1}+\ii g_{n-1}\Big) x^{a_1}  d M_{\e_{[2}}^{a_2} (dx^a)^{3\rightarrow n}(K^a)^{n+1\rightarrow p}\d_{\e_{1]}} K_{a_{1\rightarrow p}} \marginnote{(4)}% line 4 from notes
  \no \\
 &+2\sum_{n=1}^p\Big(n e_n+(n-p-1)e_{n-1}\Big)  d M_{\e_{[2}}^{a_1} (dx^a)^{2\rightarrow n}(K^a)^{n+1\rightarrow p}(\bar{\e}_{1]}\G_{a_{1\rightarrow p}}\q) \marginnote{(5)}% line 1 from notes
  \no \end{align}
  \begin{align}
  & +2\sum_{n=1}^p\Big(f_n-\ii (-1)^n k_{n-1}\Big)  M_{\e_{[2}}^{a_1} (dx^a)^{2\rightarrow n}(K^a)^{n+1\rightarrow p}
   \d_{\e_{1]}} K_{a_{1\rightarrow p}} \marginnote{(6)} % line 3 from notes
   \no \\
    &+2\sum_{n=2}^p\ii(-1)^{n+1} h_{n-1} M_{\e_{[2}}^{a_1} M_{\e_{1]}}^{a_2} (dx^a)^{3\rightarrow n}(K^a)^{n+1\rightarrow p} d K_{a_{1\rightarrow p}} \marginnote{(7)} % line 7 from notes
  \no \\
  &+2\sum_{n=2}^p\Big(-\ii g_{n-1}-\ii(n-1)(-1)^n k_{n-1}+\ii (n-p-1)(-1)^n k_{n-2}\Big)\no \\ &\hspace{6cm}\times M_{\e_{[2}}^{a_1} d M_{\e_{1]}}^{a_2} (dx^a)^{3\rightarrow n}(K^a)^{n+1\rightarrow p} K_{a_{1\rightarrow p}}  \marginnote{(8)}% line 5 from notes
  \no \\
  &+2\sum_{n=3}^p\Big(\ii(-1)^n (n-2)h_{n-1}-\ii(-1)^n (n-p-1) h_{n-2}\Big)\no \\ &\hspace{6cm}\times x^{a_1} M_{\e_{[2}}^{a_2} d M_{\e_{1]}}^{a_3} (dx^a)^{4\rightarrow n}(K^a)^{n+1\rightarrow p} d K_{a_{1\rightarrow p}} \, \no\marginnote{(9)}% line 8 from notes
\end{align}
where the square brackets indicate anti-symmetrisation.  For clarity it should be noted that the above expression is the result of a direct calculation starting with \eqref{varAe}, and we have merely reordered the various factors, performed a change of variables in some of the summations and collected terms together.  No gamma matrix identities were used.  For ease of reference we have labelled each line of the above result.  Note that all terms that appear in this result come in one of two types: type 1 terms -- lines (1), (2) and (3) --  containing $(\bar{\e}_{1}\G^{a}\e_2)$ or $(\bar{\e}_{1}\G_{a_{1\rightarrow p}}\e_2)$ factors; and type 2 terms -- lines (4) to (9) -- where $\e_1$ and $\e_2$ are not both contracted with $\G_{a_{1\rightarrow p}}$ or the same gamma matrix.

So that the supersymmetry transformation $A^{(p)}$ provides us with a realisation of an enlarged supersymmetry algebra, we must eliminate all terms of type 2 -- these cannot be interpreted as coming from a commutator $[ \epsilon_1{}^{\alpha} Q_{\alpha},  \epsilon_2{}^{\beta} Q_{\beta}]$.  This involves fixing some of the $3p-1$ arbitrary integration constants.  The procedure used for fixing the constants is quite involved due to the fact that many (but not all) of the type 1 and type 2 terms are related through the gamma matrix identities \eqref{gammaid1}.

To begin this process we start by listing all of the gamma matrix identities relating the type 1 and type 2 terms, keeping in mind that the supersymmetric parameters $\e_1$ and $\e_2$ must appear in an antisymmetric combination.  These are:
\begin{multline}
  (\ba{\q}\G^{a_1} d \q)(\ba{\e}_1\G_{a_{1\rightarrow p}}\e_2)+(\ba{\e}_1\G^{a_1}\e_2)(\ba{\q}\G_{a_{1\rightarrow p}} d \q)\\+
  2(\ba{\e}_{[1}\G^{a_1} d \q)(\ba{\e}_{2]}\G_{a_{1\rightarrow p}}\q)-2(\ba{\e}_{[1}\G^{a_1} \q)(\ba{\e}_{2]}\G_{a_{1\rightarrow p}}d\q)=0\,, \label{gammaid3}
\end{multline}
\begin{multline}
2(\ba{\q}\G^{a_1} d \q)(\ba{\e}_{[1}\G^{a_2} \q)(\ba{\e}_{2]}\G_{a_{1\rightarrow p}}d\q)+(\ba{\e}_{[1}\G^{a_1}\q)(\ba{\e}_{2]}\G^{a_2} \q)(d\ba{\q}\G_{a_{1\rightarrow p}}d\q)\\-2 (\ba{\e}_{[1}\G^{a_1}\q)(\ba{\e}_{2]}\G^{a_2} d\q)(\ba{\q}\G_{a_{1\rightarrow p}}d\q)+(\ba{\e}_{[1}\G^{a_1}\q)(d\ba{\q}\G^{a_2}d \q)(\ba{\e}_{2]}\G_{a_{1\rightarrow p}}d\q)=0\,, \label{gammaid4}
\end{multline}
and
\begin{multline}
  (d\ba{\q}\G^{a_1} d \q)(\ba{\e}_1\G_{a_{1\rightarrow p}}\e_2)+(\ba{\e}_1\G^{a_1}\e_2)(d\ba{\q}\G_{a_{1\rightarrow p}} d \q)-
  4(\ba{\e}_{[1}\G^{a_1} d \q)(\ba{\e}_{2]}\G_{a_{1\rightarrow p}}d\q)=0\,. \label{gammaid5}
\end{multline}

By carefully examining these identities and the terms appearing in our expression for  $[\d_{\e_2},\d_{\e_1}] A^{(p)}$, we notice that none of the identities may be used to completely remove the type 2 terms appearing on lines (7) or (8) in favour of type 1 terms. That is, any application of an identity to the terms appearing in lines (7) or (8) will necessarily result in the appearance of other type 2 terms.  We are therefore required to set the coefficients of these terms to zero if we wish to fully eliminate all type 2 terms. Explicitly this means setting
      \begin{align}
          h_n&=0 \qquad n=1,\ldots, p-1 \label{h2}\\
            g_{n}+n(-1)^{n+1} k_n+(n-p)(-1)^{n} k_{n-1} &=0 \qquad n=1,\ldots, p-1\,.
      \end{align}
These, along with the expressions for $g_n$ and $k_n$, \eqref{g} and \eqref{k}, lead to the recurrence relation:
\begin{align}
  e_{n+1}=\frac{1}{2p-n}\left((p-n-1)e_n+\frac{p!}{(p-n)!(n+1)!}\right)  \qquad  n=0,\ldots, p-2\,. \label{rre}
\end{align}
Thus we have now fixed  $2(p-1)$ of the $3p-1$ arbitrary constants appearing in our expression for $\d_\e A^{(p)}$.  More explicitly, the constants $h_1, h_2, \ldots, h_{p-1}, e_1, e_2, \ldots, e_{p-1}\,$ are now fixed and only $b_1, b_2, \ldots, b_p$ and  $e_p\,$ remain arbitrary.  Notice that the vanishing of all of the constants $h_n$ also ensures that all type 2 terms appearing on line (9) vanish.

Having fixed these constants we now turn our attention to the type 2 terms appearing on lines (5) and (6). The coefficients for $n=1,2 \ldots, p-1$ appearing on line (5), and the coefficients for $n=1,2 \ldots, p$  appearing on line (6) are already fixed, since they are completely independent of the remaining arbitrary constants.  It follows that the type 2 terms appearing on lines (5) and (6) can only be eliminated through the gamma matrix identity \eqref{gammaid3}, and this is only possible provided the following constraint is satisfied:
  \begin{align}
  n e_n+(n-p-1)e_{n-1}=(-1)^p(f_n-\ii (-1)^n k_{n-1}) \qquad n=1,2,\ldots p\,.
\end{align}
One can readily check that this expression already holds for $n=1,2,\ldots, p-1$ due to \eqref{f}, \eqref{k} and \eqref{rre}.  We must further insist that it holds for $n=p$ (otherwise additional type 2 terms will remain or be generated), thus fixing the constant $e_p$.  This extends the validity of the recurrence relation \eqref{rre} to include the case $n=p-1$.

The only remaining type 2 terms occur on line (4), the coefficients of which are now completely fixed.  These terms must therefore be recast into type 1 terms by usage of the identity \eqref{gammaid5}.

After some simplification, all of this leaves us with the following result:
\begin{multline}
  [\d_{\e_2},\d_{\e_1}] A^{(p)}=d\Bigg[\sum_{n=1}^{p}w_n x^{a_1}(dx^a)^{2\rightarrow n}(K^a)^{n+1\rightarrow p}\Bigg](\ba{\e}_1\G_{a_{1\rightarrow p}}\e_2) \\-2(\ba{\e}_1\G^{a_1}\e_2)\left[\sum_{n=1}^{p} u_n (dx^a)^{2\rightarrow n}(K^a)^{n+1\rightarrow p} K_{a_{1\rightarrow p}}
   +\sum_{n=2}^{p}v_n x^{a_2}(dx^a)^{3\rightarrow n}(K^a)^{n+1\rightarrow p}d K_{a_{1\rightarrow p}}\right]   \label{ddA}
\end{multline}
with
\begin{align}
  u_n&=(-1)^p\frac{\ii }{2}\left(\frac{3 p!}{n!(p-n+1)!} - (2 p - 2 n + 1) e_n-3 e_{n-1} -2 (-1)^p b_n\right) \qquad  n=1,\ldots, p\,, \\
  v_n&=\frac{\ii}{4}\left(\frac{p!}{n!(p-n+1)!}+(2 p - 2 n + 1)(n-1) e_n  -(2 p - 2 n + 3)(p-n+2) e_{n-1}\right) \no \\ & \hspace{11cm}n=2,\ldots, p\,, \\
  w_n&=\left(\frac{3p+2(p-n)^2+1}{2p-n}\right)e_n -\frac{p!}{(2p-n) n!(p-n)!} \qquad  n =1,\ldots, p\,, \label{w}
\end{align}
where $b_1, b_2, \ldots, b_p$ are arbitrary constants and the $e_n$ are given by the recurrence relation:
\begin{align}
 e_0&=\frac{1}{(p+1)(2p+1)}\label{e0}\, \\
  e_{n+1}&=\frac{1}{2p-n}\left((p-n-1)e_n+\frac{p!}{(p-n)!(n+1)!}\right)   \qquad  n=0,\ldots, p-1\,. \label{e2}
\end{align}

\section{Computing the algebra}\label{sec:computingthealgebra}
\setcounter{equation}{0}

If we use $\d_\e=\e^\a Q_\a$ and require that \eqref{ddA} be consistent with an enlargement of the supersymmetry algebra of the form \cite{de Azcarraga:1989gm}
\begin{equation}
  \left\{Q_\a,Q_\b\right\}=-2(C\G^a)_{\a\b}P_a+ (C\G_{a_{1\rightarrow p}})_{\a\b}Z_{0}^{a_{1\rightarrow p}} \label{A1}
\end{equation}
where $Z_{0}^{a_{1\rightarrow p}}=Z_0^{a_1 a_2\cdots a_p}$ is a bosonic charge\footnote{The subscript  ``0'' appears on this charge since we anticipate that the entire algebra will involve $p+1$ charges.  The subscript is chosen so that it `counts' the number of spinor indices carried by the change.} completely antisymmetric in its indices, we infer that as on operator relation in ``$(x,\q,A)$ space'',
\begin{align}
  P_{a_1} A^{(p)}&=  \sum_{n=1}^{p}u_n(dx^a)^{2\rightarrow n}(K^a)^{n+1\rightarrow p} K_{a_{1\rightarrow p}} + \sum_{n=2}^{p}v_n x^{a_2}(dx^a)^{3\rightarrow n}(K^a)^{n+1\rightarrow p}d K_{a_{1\rightarrow p}} \label{PA}
  \intertext{and}
 Z_{0}^{a_{1\rightarrow p}}A^{(p)} &= d\Bigg[\sum_{n=1}^{p}w_n x^{[a_1}(dx^a)^{2\rightarrow n}(K^{a]})^{n+1\rightarrow p}\Bigg]\,. \label{Z0A}
\end{align}

We are now in a position to systematically determine the rest of the algebra. For example, by applying\footnote{Compatibility of  $[\d_{\e_2},\d_{\e_1}]x^a = 2\ii(\ba{\e}_1\G^{a}\e_2)$ with the algebra requires $P_a x^b = -\ii \d_a^b$.} $P_a$ to \eqref{varAe} with its constants subject to \eqref{f}-\eqref{k}, \eqref{h2}, \eqref{e0} and \eqref{e2}, and by computing the supersymmetry transformation of \eqref{PA} and using $\d_\e=\e^\a Q_\a$, we find, after extensive simplification and employing \eqref{sgi2}, that
\begin{multline}
  [Q_\a,P_b]A^{(p)}=-\frac{1}{2}(C\G_{ba_{1\rightarrow p-1}})_{\a\b} d\Bigg[\sum_{n=1}^{p}\ii w_n (dx^{[a})^{1\rightarrow n-1}(K^{a]})^{n\rightarrow p-1}\q^\b   \\
 +\sum_{n=2}^{p}z_n x^{[a_1}(dx^a)^{2\rightarrow n-1}(K^{a]})^{n\rightarrow p-1}d \q^\b \Bigg] \,
\end{multline}
where
\begin{align}
  z_n=(-1)^{p-1}\Big(\ii w_n(n-1)-\ii w_{n-1}(p-n+1)\Big) \qquad \qquad n=2,\ldots, p \,.
\end{align}
This leads us to introduce a fermionic charge $Z_{1}^{a_{2\rightarrow p}\b}$, antisymmetric in its Lorentz indices, defined\footnote{There is always the freedom to re-scale any of the charges we introduce.  Throughout this paper we choose scales so that our algebra is consistent with our previous work.}  via
\begin{align}
  [Q_\a,P_b]=(C\G_{ba_{2\rightarrow p}})_{\a\b} Z_{1}^{a_{2\rightarrow p}\b}
\end{align}
where
\begin{multline}
 Z_{1}^{a_{1\rightarrow p-1}\b}A^{(p)}=-\frac{1}{2} d\Bigg[\sum_{n=1}^{p}\ii w_n (dx^a)^{1\rightarrow n-1}(K^{a]})^{n\rightarrow p-1}\q^\b \\
 +\sum_{n=2}^{p}z_n x^{[a_1}(dx^a)^{2\rightarrow n-1}(K^{a]})^{n\rightarrow p-1}d \q^\b \Bigg]\,. \label{Z1A}
\end{multline}

Similarly, computing the supersymmetry transformation of \eqref{Z0A}, using $\d_\e=\e^\a Q_\a$, and assuming that $x^a$ and $\q^\a$ are inert under the action of the charges\footnote{An assumption which is extended to all of the $Z$ charges in what follows.}  we find
\begin{align}
[Q_\a,Z_0^{ba_{2\rightarrow p}}]A^{(p)}=-2(C\G^{[b})_{\a\b} Z_{1}^{a_{2\rightarrow p}]\b}A^{(p)}\,.
\end{align}

By computing the supersymmetry transformation of \eqref{Z1A}, we find
\begin{multline}
[Q_\a, \,Z_1^{b a_{1\rightarrow p-2}\b}]A^{(p)}=-\frac12 (C\G^{[b})_{\l\g}
d\Bigg[\sum_{n=1}^{p-1}(-1)^p w_n (dx^a)^{1\rightarrow n-1}(K^{a]})^{n\rightarrow p-2}\q^\l d \q^\g\Bigg]\d_\a^\b  \\
 -\frac12(C\G^{[b})_{\a\g}d\Bigg[\sum_{n=2}^{p}2\ii z_n (dx^a)^{1\rightarrow n-2}(K^{a]})^{n-1\rightarrow p-2}\q^{(\g} d\q^{\b)} \Bigg.  \\
+\sum_{n=3}^{p}(-1)^p\Big(\ii z_n(n-2)-\ii z_{n-1}(p-n+1)\Big) x^{[a_1}(dx^a)^{2\rightarrow n-2}(K^{a]})^{n-1\rightarrow p-2}d \q^\g d \q^\b\Bigg]\,.\label{QZ1}
\end{multline}

We also find that the application of $P_a$ to \eqref{Z0A} yields
\begin{align}
[P_b, Z_0^{cd a_{1\rightarrow p-2}}]A^{(p)}=-\d_b^{[c}(C\G^{d})_{\a\b}
d\Bigg[\sum_{n=1}^{p-1}(-1)^p w_n (dx^a)^{1\rightarrow n-1}(K^{a]})^{n\rightarrow p-2}\q^\a d \q^\b\Bigg]\,.
\end{align}
Thus we are lead to introduce a bosonic charge $Z_{2}^{a_{1\rightarrow p-2}\a\b}$, antisymmetric in its Lorentz indices and symmetric in its spinor indices, such that
\begin{align}
[Q_\a, Z_1^{b a_{1\rightarrow p-2}\b}] =-\frac12 (C\G^{[b})_{\l\g} Z_{2}^{a_{1\rightarrow p-2}]\l\g}\d_\a^\b -4(C\G^{[b})_{\a\g}Z_{2}^{a_{1\rightarrow p-2}]\g\b}\,
\end{align}
and
\begin{align}
[P_b, Z_0^{cd a_{1\rightarrow p-2}}]=-\d_b^{[c}(C\G^{d})_{\a\b}Z_{2}^{a_{1\rightarrow p-2}]\a\b}\,
\end{align}
where
\begin{multline}
 Z_{2}^{a_{1\rightarrow p-2}\a\b} A^{(p)}=\frac18 d\Bigg[\sum_{n=2}^{p}2\ii z_n (dx^{[a})^{1\rightarrow n-2}(K^{a]})^{n-1\rightarrow p-2}\q^{(\a} d\q^{\b)} \Bigg. \\
+\sum_{n=3}^{p}(-1)^p\Big(\ii z_n(n-2)-\ii z_{n-1}(p-n+1)\Big) x^{[a_1}(dx^a)^{2\rightarrow n-2}(K^{a]})^{n-1\rightarrow p-2}d \q^\a d \q^\b\Bigg]\,. \label{Z2A}
\end{multline}
The only non-trivial step here is establishing that, for $Z_{2}^{a_{1\rightarrow p-2}\a\b} A^{(p)}$ as given in \eqref{Z2A}, the following holds:
\begin{align}
  (C\G^{[b})_{\a\b}Z_{2}^{a_{1\rightarrow p-2}]\a\b} A^{(p)} = (C\G^{[b})_{\a\b} d\Bigg[\sum_{n=1}^{p-1}(-1)^p w_n (dx^a)^{1\rightarrow n-1}(K^{a]})^{n\rightarrow p-2}\q^\a d \q^\b\Bigg]\,\label{colid}
\end{align}
which requires recognising that for $3 \leq n \leq p$
\begin{align}
  d\bigg[ x^{[a_1}(dx^a)^{2\rightarrow n-2}(K^{a})^{n-1\rightarrow p-2}d K^{b]}\bigg]=\frac{(-1)^{p}}{p-n+1}d\bigg[(dx^{[a})^{1\rightarrow n-2}(K^{a})^{n-1\rightarrow p-2} K^{b]}\bigg] \label{colid1}
\end{align}
and showing that both of the following are satisfied
\begin{align}
  \ii (-1)^{p-1}8 w_1&=2z_2 \label{wz1}\\
  \ii (-1)^{p-1} 8 w_{n-1}& = \frac{2p-n}{p-n+1}z_n- z_{n-1} \qquad  3 \leq n \leq p\,.\label{wz2}
\end{align}

The application of $P_a$ to expression \eqref{PA} immediately leads to the following expression:
\begin{align}
  [P_b,P_c] A^{(p)}&= \sum_{n=2}^{p}2 \ii v_n (dx^a)^{1\rightarrow n-2}(K^a)^{n-1\rightarrow p-2}d K_{bca_{1\rightarrow p-2}} \,.
\end{align}
By utilising \eqref{sgi} and \eqref{sgi1}, and by establishing
 \begin{align}
  2\ii v_2&=\frac{\ii }{4}(-1)^p(p-1)z_2 \\
  2\ii v_n& = \frac{\ii}{8}(-1)^p\big((2p-n)z_n- (p-n+1)z_{n-1}\big) \qquad  3 \leq n \leq p\,,
\end{align}
it can be demonstrated that $Z_{2}^{a_{1\rightarrow p-2}\a\b} A^{(p)}$ as given in \eqref{Z2A} satisfies
\begin{align}
 (C\G_{bca_{1\rightarrow p-2}})_{\a\b} Z_{2}^{a_{1\rightarrow p-2}\a\b}A^{(p)}= \sum_{n=2}^{p}2 \ii v_n (dx^a)^{1\rightarrow n-2}(K^a)^{n-1\rightarrow p-2}d K_{bca_{1\rightarrow p-2}} \,,
\end{align}
from which it immediately follows that
\begin{align}
  [P_b,P_c] A^{(p)}&= (C\G_{bca_{3\rightarrow p}})_{\a\b} Z_{2}^{a_{3\rightarrow p}\a\b}A^{(p)}\,.
\end{align}

Determining the remainder of the algebra is relatively straight forward, although calculational intensive. In what follows we present the broad outline of the calculation, giving only the important results.

By scrutinising what has been done so far, a clear pattern presents itself.  When we repeat the procedures given above a set of charges $Z_k^{a_{1\rightarrow p-k}\a_{1\rightarrow k}}$ emerges, where $1 \leq k \leq p$.  The charge $Z_k^{a_{1\rightarrow p-k}\a_{1\rightarrow k}}$ carries a parity of $k$ mod 2, is antisymmetric in its $p-k$ Lorentz indices and symmetric in its $k$ spinor indices.  Based on the pattern and the above experience, we establish the remainder of the algebra by postulating that the action of these charges on $A^{(p)}$ is of the form
\begin{multline}
 Z_{k}^{a_{1\rightarrow p-k}\a_{1\rightarrow k}}A^{(p)}=J_k d\Bigg[\sum_{n=k}^{p}L_{k,n} (dx^{[a})^{1\rightarrow n-k}(K^{a]})^{n-k+1\rightarrow p-k}\q^{(\a_1}(d \q^{\a)})^{2\rightarrow k} \Bigg.\\
 + \Bigg. \sum_{n=k+1}^{p}M_{k,n} x^{[a_1}(dx^a)^{2\rightarrow n-k}(K^{a]})^{n-k+1\rightarrow p-k} (d \q^{\a})^{1\rightarrow k} \Bigg]\, \qquad 1 \leq k \leq p \label{ZkA}
\end{multline}
with
\begin{align}
 M_{k,n}= \frac{(-1)^{p-k}}{k}\Big(L_{k,n}(n-k)-L_{k,n-1}(p-n+1)\Big)\qquad  k+1\leq n\leq p\,,\, 1\leq k\leq p-1
\end{align}
and satisfy the following (anti)commutation relations\footnote{We adopt the graded commutator notation $\gc{X}{Y}=XY-(-1)^{\ve(X)\ve(Y)} YX\,,$ where $\ve(X)$, for example, denotes the parity of $X$.}
\begin{align}
 \gc{P_b}{Z_k^{cd a_{3\rightarrow p-k}\a_{1\rightarrow k}}}\!A^{(p)}&=-\d_b^{[c}(C\G^{d})_{\l\g} Z_{k+2}^{a_{3\rightarrow p-k}]\l\g\a_{1\rightarrow k}}A^{(p)} && 0 \leq k \leq p-2 \label{A2}\\ \gc{Q_\b}{Z_k^{ba_{2\rightarrow p-k}\m \a_{2\rightarrow k}}}\!A^{(p)}&=s_k(C\G^{[b})_{\l\g} Z_{k+1}^{a_{2\rightarrow p-k}]\l\g(\a_{2\rightarrow k}}\d_\b^{\m)}A^{(p)}\no \\ & \qquad  +t_{k}(C\G^{[b})_{\b\g} Z_{k+1}^{a_{2\rightarrow p-k}]\g\m\a_{2\rightarrow k}}A^{(p)} && 1 \leq k \leq p-1 \label{A3}
\end{align}
where $J_k$, $L_{k,n}$,  $s_{k}$ and $t_k$ are a collection of constants.  Our previous expressions certainly take the above form with
\begin{align}
  J_1=-\frac12\,,\quad  J_2=\frac18\,,\quad L_{1,n}=\ii w_n\,,\quad L_{2,n}=2 \ii z_n\,, \quad s_1=-\frac12\,,\quad t_1=-4\,. \label{ff}
\end{align}
We can now proceed inductively and establish that the  postulated actions above are indeed correct where:
\begin{align}
  L_{k+1,n}&=\ii(k+1)M_{k,n}  &  &k+1\leq n\leq p\,,\, 1\leq k\leq p-1 \label{L}\\
  s_k&=\frac{k J_{k}}{J_{k-1}}\,, & & 2\leq k\leq p-1 \label{s}\\
  t_k&=\frac{J_k}{J_{k+1}}\,,    & & 1\leq k\leq p-1\,   \label{t}
\end{align}
and
\begin{align}
 (-1)^{p-k}\left(\frac{t_k}{s_k}\right)L_{k,k}&=\frac{2p-k-1}{p-k}M_{k,k+1} & &1\leq k\leq p-1 \label{KM}\\
 (-1)^{p-k}\left(\frac{t_k}{s_k}\right)L_{k,n-1} &=\frac{2p-n}{p-n+1}M_{k,n}-M_{k,n-1} & &k+2\leq n\leq p\,,\, 1\leq k\leq p-2\,. \label{KM1}
\end{align}
The last two expressions are the generalisations of equations \eqref{wz1} and \eqref{wz2}, many of which may be viewed as consistency conditions.  As with our previous calculations, the only non-trivial step in the inductive procedure is in establishing that a generalization of identity \eqref{colid} holds.  More specifically, we are required to demonstrate
\begin{multline}
  (C\G^{[b})_{\l\g} Z_{k+1}^{a_{1\rightarrow p-k-1}\l\g \a_{1\rightarrow k-1}}A^{(p)} = \frac{J_k}{s_k}(C\G^{[b})_{\l\g} \\\times d\Bigg[\sum_{n=k+1}^{p}(-1)^{p-k}\ii L_{k,n-1} (dx^a)^{1\rightarrow n-k-1}(K^{a]})^{n-k\rightarrow p-k-1}\q^\l d \q^\g (d \q^{\a})^{1\rightarrow k-1}\Bigg]\,,
\end{multline}
which involves using identities \eqref{KM} and \eqref{KM1} and the following generalisation of identity \eqref{colid1}
\begin{multline}
d\bigg[ x^{[a_1}(dx^a)^{2\rightarrow n-k-1}(K^{a})^{n-k\rightarrow p-k-1}d K^{b]}\bigg]\\=\frac{(-1)^{p-k-1}}{p-n+1}d\bigg[(dx^{[a})^{1\rightarrow n-k-1}(K^{a})^{n-k\rightarrow p-k-1} K^{b]}\bigg] \qquad  k+2\leq n\leq p\,,\, 1\leq k\leq p-2\,,
\end{multline}
in addition to the identity\footnote{Note the symmetrisation of the spinor indices.}
\begin{multline}
(C\G^{[b})_{\l\g} d\Bigg[(dx^a)^{1\rightarrow n-k-1}(K^{a]})^{n-k\rightarrow p-k-1}\q^{(\l} d \q^\g (d \q^{\a)})^{1\rightarrow k-1}\Bigg]
=\frac{2 p -2 n + k + 1}{(p - n + 1) (k + 1)}\\\times(C\G^{[b})_{\l\g} d\Bigg[(dx^a)^{1\rightarrow n-k-1}(K^{a]})^{n-k\rightarrow p-k-1}\q^{\l} d \q^\g (d \q^{\a})^{1\rightarrow k-1}\Bigg]\qquad  k+1\leq n\leq p\,,\, 1\leq k\leq p-1\,.
\end{multline}

At this stage our task is effectively complete since we have determined the entire algebra for all valid $p$, and we have obtained expressions for the action of the charges on $A^{(p)}$.  Using a computer it is a trivial matter to implement the recursion relations to determine the constants $L_{k,n}$, $J_k$, $t_k$ and $s_k$ and check that the consistency conditions hold. Quite remarkably, in doing so, we find that although the constants $L_{k,n}$ depend on the value of $p$, the constants $J_k$, $t_k$ and $s_k$ are completely $p$ independent.  Expressions \eqref{ff} provide an example of this.

Although we are unable to directly establish closed form (i.e. non-recursive) expressions for the constants $L_{k,n}$, we are able to do so for $J_k$, $t_k$ and $s_k$.  This is achieved by exploiting the Jacobi identities.  In particular we focus on the following set:
\begin{align}
  \gc{Q_\b}{\gc{Q_\n}{Z_k}}+\gc{Q_\n}{\gc{Q_\b}{Z_k}}+\gc{Z_k}{\gc{Q_\b}{Q_\n}}=0 \label{genJI}
\end{align}
where $Z_k$ is a charge with spacetime and spinor indices suppressed.

Using \eqref{A1}, \eqref{A2} and \eqref{A3}, we find that the Jacobi identities  \eqref{genJI} with $1\leq k\leq p-2$ become
\begin{align}
& \left(\frac{2t_k s_{k+1}}{k+1} -2 \right) (C\G^{[b})_{\n\b}(C\G^{c})_{\l\g}Z_{k+2}^{a_{3\rightarrow p-k}]\l\g\m\a_{2\rightarrow k}}\no\\
&-2\left(\frac{k t_k s_{k+1}}{k+1} +\frac{2 s_k s_{k+1}}{k+1}-t_{k+1}s_k \right) (C\G^{[b})_{\l\g}(C\G^{c})_{\r(\b}Z_{k+2}^{a_{3\rightarrow p-k}]\l\g\r(\a_{2\rightarrow k}}\d_{\n)}^{\m)}=0\,,
\end{align}
where the coefficients appearing on each line should vanish identically.  The coefficient on the first line is already known to vanish by combining \eqref{s} and \eqref{t}.  The vanishing of the coefficient on the second line provides us with a new relationship between $s_t$ and $t_k$, which can be rearranged to become:
\begin{align}
  \frac{t_{k+1}}{s_{k+1}}=\frac{k}{k+1}\frac{t_k}{s_k}+ \frac{2}{k+1}\ \qquad \qquad 1\leq k\leq p-2\,.
\end{align}
This expression is clearly a recurrence relation in the ratio $t_k/s_k$, which, together with the knowledge $t_1/s_1=8$ from \eqref{ff}, yields the following solution
\begin{align}
 \frac{t_k}{s_k}=\frac{6 + 2 k}{k} \qquad \qquad 1 \leq k \leq p-1\,.
\end{align}
Together with \eqref{s},  \eqref{t} and \eqref{ff}, it then follows that
\begin{align}
  J_1=-\frac12\,,\quad  J_2=\frac18\,,\qquad \qquad  J_{k+1} =\frac{J_{k-1}}{6+2k}\,\qquad 2 \leq k \leq p-1
\end{align}
which is readily solved to give:
\begin{align}
  J_k=\frac{4 \big(1 + (-1)^k\big) + 3 \sqrt{\p}\big(-1 + (-1)^k\big)}{2^{k+3}\,\G\!\left(2 + \tfrac{k}{2}\right)}\,\qquad 1 \leq k \leq p  \label{J}
\end{align}
where here $\G$ is the Euler gamma function.  Using this solution along with \eqref{s} and \eqref{t}, we find the constants $s_k$ and $t_k$, with $1 \leq k \leq p-1$, are given by
\begin{align}
  s_k&=\frac{k \,t_k }{6+2 k}\,, & t_k &=-\frac{2 \left(4 (1 + (-1)^k) + 3\sqrt{\p} (-1 + (-1)^k) \right) \G\!\left(\tfrac{5 + k}{2}\right)}{\left(4 (-1 + (-1)^k) + 3\sqrt{\p} (1 + (-1)^k) \right) \G\!\left(\tfrac{4 + k}{2}\right)}\,.
\end{align}

\section{The general algebra for all valid $p$}\label{sec:thealgebra}
\setcounter{equation}{0}

Here we summarize our final results regarding the enlarged superalgebras underlying $p$-brane actions. Our results hold for any combination of spacetime dimension and $p$ for which the identity (\ref{gammaid1}) is valid, namely for the ``$p$-brane scan'' in Table \ref{table}.

A total of $p+1$ additional charges $Z$ make an appearance in the algebra.  The notation $Z_k^{a_{1\rightarrow p-k}\a_{1\rightarrow k}}$ indicates that the charge is carrying $p-k$ antisymmetrized Lorentz indices $a_i$, $k$ symmetrized spinor indices $\a_i$, and a parity of $k$ mod 2.

The algebra is\footnote{Note that there are no generators $Z_k$ for $k>p$ and so the $p = 1$ algebra is recovered by formally setting the $Z_2$ generator to zero.}:
\begin{align}
  \gc{Q_\a}{Q_\b}&=-2(C\G^a)_{\a\b}P_a+ (C\G_{a_{1\rightarrow p}})_{\a\b}Z_{0}^{a_{1\rightarrow p}} \no\\
  \gc{Q_\a}{P_b}&=(C\G_{ba_{2\rightarrow p}})_{\a\b} Z_{1}^{a_{2\rightarrow p}\b} \no\\
  \gc{P_b}{P_c}&=(C\G_{bca_{3\rightarrow p}})_{\a\b} Z_{2}^{a_{3\rightarrow p}\a\b}\no\\
  \gc{Q_\a}{Z_0^{ba_{2\rightarrow p}}}&=-2(C\G^{[b})_{\a\b} Z_{1}^{a_{2\rightarrow p}]\b}\no\\
  \gc{P_b}{Z_k^{cd a_{3\rightarrow p-k}\a_{1\rightarrow k}}}&=-\d_b^{[c}(C\G^{d})_{\l\g} Z_{k+2}^{a_{3\rightarrow p-k}]\l\g\a_{1\rightarrow k}} && 0 \leq k \leq p-2 \no\\
  \gc{Q_\b}{Z_k^{ba_{2\rightarrow p-k}\m \a_{2\rightarrow k}}}&=s_k(C\G^{[b})_{\l\g} Z_{k+1}^{a_{2\rightarrow p-k}]\l\g(\a_{2\rightarrow k}}\d_\b^{\m)}\no \\ & \qquad \qquad \qquad  +t_{k}(C\G^{[b})_{\b\g} Z_{k+1}^{a_{2\rightarrow p-k}]\g\m\a_{2\rightarrow k}} && 1 \leq k \leq p-1 \no
\end{align}
where the constants $s_k$ and $t_k$, with $1 \leq k \leq p-1$, are given by
\begin{align}
  s_k&=\frac{k \,t_k }{6+2 k}\,, & t_k &=-\frac{2 \left(4 (1 + (-1)^k) + 3\sqrt{\p} (-1 + (-1)^k) \right) \G\!\left(\tfrac{5 + k}{2}\right)}{\left(4 (-1 + (-1)^k) + 3\sqrt{\p} (1 + (-1)^k) \right) \G\!\left(\tfrac{4 + k}{2}\right)}\,
\end{align}
with $\G$ the Euler gamma function.

With appropriate rescaling of generators, the general algebra given above incorporates, as special cases, all of the known algebras from $p=1$ to $p=5$, which are all separately displayed in \cite{Reimers:2005cp} (which come from \cite{Bergshoeff:1995hm} and \cite{Chryssomalakos:1999xd} for $p=1$, \cite{Bergshoeff:1995hm} for $p=2$ and $p=3$, and have been derived for the cases $p=4$ and $p=5$ from an ansatz for Maurer-Cartan equations appearing in \cite{Chryssomalakos:1999xd}).  Additionally, taking into account notational differences, the right hand sides of our expressions \eqref{Z0A} and \eqref{ZkA} -- where the constants are given by \eqref{w}-\eqref{e2}, \eqref{ff}, \eqref{L} and \eqref{J} -- reproduces the $p=1$ to $p=5$ Noether charges constructed in \cite{Reimers:2005cp} (up to a different choice of normalization). Previously, all of these results were determined on a case by case basis -- here we have given a general expression applicable to any valid value of $p$.

\section{Computing $\mathcal{F}^{(p+1)}$}\label{sec:F}
\setcounter{equation}{0}
Having determined a compact expression for the entire algebra for arbitrary $p$, we can now derive an expression for $\mathcal{F}^{(p+1)}=d A^{(p)}-b^{(p+1)}$ using a coset construction based on the enlarged superalgebra. This generalises earlier constructions for supersymmetric $p$-brane Wess-Zumino terms for specific values of $p$  \cite{Siegel:1994xr, Bergshoeff:1989ax, Bergshoeff:1995hm, Sezgin:1996cj, Chryssomalakos:1999xd}.

We begin by introducing the following coset parametrisation
\begin{align}\label{coset}
  \W(x, \q, y )=\exp\left\{\ii \left(x^a P_a+\q^\a Q_\a+\sum_{k=0}^{p} y^{k}_{a_{1\rightarrow p-k}\a_{1\rightarrow k}}  Z_{k}^{a_{1\rightarrow p-k}\a_{1\rightarrow k}} \right)\right\}\,,
\end{align}
where $(x^a, \q^\a, y^{0}_{a_{1\rightarrow p}},y^{1}_{a_{1\rightarrow p-1}\a_1},\ldots, y^{p}_{\a_{1\rightarrow p}})$ are the extended superspace coordinates.  Transformations of these coordinates are determined by considering the left action of group elements $g$ on $\W(x, \q, y )$, specifically\footnote{With respect to the $k$ index appearing on $Z_k$, $y^{k}$ and $h^{k}$, all summations will be made explicit.}:
\begin{align}\label{left}
  g= \exp\left(\ii \e^{\a}Q_{\a}\right)\,, \quad   g=\exp\left(\ii  c^{a}P_{a}\right)\,, \quad g=\exp\left(\ii h^{k}_{a_{1\rightarrow p-k}\a_{1\rightarrow k}}  Z_{k}^{a_{1\rightarrow p-k}\a_{1\rightarrow k}}\right)
\end{align}
which, respectively, correspond to supersymmetry transformations with parameter $\e^{\a}$, translations with parameter $c^{a}$, and a ``$Z_k$ transformation'' with parameter $h^{k}$ (having suppressed spacetime and spinor indices).

From the resulting extended superspace coordinate transformations, consistency with the algebra, and our earlier use of $\d_\e=\e^\a Q_\a$, we infer that as operators generating infinitesimal translations and $Z_k$ transformations,
\begin{align}
 \d_c=\ii c^a P_a \qquad \textrm{and} \qquad \d_{h^k}=\ii^{k+1}h^{k}_{a_{1\rightarrow p-k}\a_{1\rightarrow k}}  Z_{k}^{a_{1\rightarrow p-k}\a_{1\rightarrow k}}\,.
\end{align}
Additionally, it is not difficult to establish that for an infinitesimal $Z_k$ transformation with parameter $h^{k}$, the extended superspace coordinates $x^a$ and $\q^\a$ are inert and
\begin{align}
  \d_{h^k}y^{m}_{a_{1\rightarrow p-m}\a_{1\rightarrow m}}\Big| = \d_k{}^m h^{k}_{a_{1\rightarrow p-k}\a_{1\rightarrow k}}
\end{align}
with $0 \leq m\leq p$, where $ \d_k{}^m$ on the right hand side is the Kronecker delta.  Here we have introduced the notation $|$ to denote the act of setting all enlarged superspace coordinates to zero: $x^a=0$, $\q^\a=0$, and $y^m=0$ for $0 \leq m\leq p$. For later use we note that \eqref{Z0A} and \eqref{ZkA} now yield
\begin{align}
  \d_{h^0}A^{(p)}\Big|&=\ii h^{0}_{a_{1\rightarrow p}}Z_{0}^{a_{1\rightarrow p}}A^{(p)}\Big|=\ii w_p h^{0}_{a_{1\rightarrow p}} (dx^a)^{1\rightarrow p} \label{varAp0} \\
  \d_{h^k}A^{(p)}\Big|&=\ii^{k+1}  h^{k}_{a_{1\rightarrow p-k}\a_{1\rightarrow k}} Z_{k}^{a_{1\rightarrow p-k}\a_{1\rightarrow k}} A^{(p)}\Big|
 \no \\ &=\ii^{k+1}J_k\left[(-1)^{p-k}L_{k,p}+M_{k,p}\right] h^{k}_{a_{1\rightarrow p-k}\a_{1\rightarrow k}} (dx^a)^{1\rightarrow p-k}(d \q^{\a})^{1\rightarrow k} \qquad 1\leq k\leq p-1 \label{varApk}\\
 \d_{h^p}A^{(p)}\Big|&=\ii^{p+1}  h^{p}_{\a_{1\rightarrow p}} Z_{p}^{\a_{1\rightarrow p}} A^{(p)}\Big|
 =\ii^{p+1}J_p L_{p,p} h^{p}_{\a_{1\rightarrow p}}(d \q^{\a})^{1\rightarrow p} \label{varApp}\,.
\end{align}

By examining the form of the infinitesimal enlarged superspace coordinate transformations which result from \eqref{left}, and given the known effect of the generators $P_a$ and $Z_{k}$ on $A^{(p)}$, \eqref{Z0A} and \eqref{ZkA}, and having only a small collection of possible forms from which it can be constructed, we can readily infer that, when written explicitly in terms of the enlarged superspace coordinates, $A^{(p)}$ must have the form
\begin{align}
A^{(p)}=\sum_{m=0}^{p}\sum_{n=0}^{p-m} a_{n,m}y^{m}_{a_{1\rightarrow p-m}\a_{1\rightarrow m}} (dx^a)^{1\rightarrow n}(K^a)^{n+1\rightarrow p-m}(d \q^{\a})^{1\rightarrow m}+ \cdots
\end{align}
Here $a_{n,m}$ is a collection of constants and the ellipsis indicate additional terms which do not contribute to the following expressions :
\begin{align}\label{varAp}
\d_{h^k} A^{(p)}\Big|= a_{p-k,k}h^{k}_{a_{1\rightarrow p-k}\a_{1\rightarrow k}} (dx^a)^{1\rightarrow p-k}(d \q^{\a})^{1\rightarrow k}\,,
\end{align}
and
\begin{align}
d A^{(p)}\Big|= \sum_{m=0}^{p} a_{p-m,m}dy^{m}_{a_{1\rightarrow p-m}\a_{1\rightarrow m}} (dx^a)^{1\rightarrow p-m}(d \q^{\a})^{1\rightarrow m}\,.
\end{align}

Comparing \eqref{varAp} with \eqref{varAp0}-\eqref{varApp} we can read off the constants $a_{p-k,k}$:
\begin{align}
  a_{p,0}&=\ii w_p \label{ap0}\\
  a_{p-k,k}&=\frac{\ii^{k+1}(-1)^{p-k}J_k}{k}\big(p L_{k,p}-L_{k,p-1}\big) \qquad 1 \leq k \leq p-1 \label{apk}\\
  a_{0,p}&=\ii^{p+1}J_p L_{p,p} \label{app}\,.
\end{align}

Returning to the coset parametrisation \eqref{coset}, we note that corresponding Maurer-Cartan forms are
\begin{align}
 \W(x, \q, y)^{-1} d \W(x, \q, y) =  \ii \left(E^a P_a+E^\a Q_\a+ \sum_{k=0}^{p} E^{k}_{a_{1\rightarrow p-k}\a_{1\rightarrow k}}  Z_{k}^{a_{1\rightarrow p-k}\a_{1\rightarrow k}}  \right)\,
\end{align}
and it is not difficult to establish that
\begin{align}
  E^a\Big|=dx^a\,,\qquad E^\a\Big| =d \q^{\a} \,,\qquad E^{k}\Big|=dy^{k}_{a_{1\rightarrow p-k}\a_{1\rightarrow k}}\,.
\end{align}
Since $\mathcal{F}^{(p+1)}=d A^{(p)}-b^{(p+1)}$, it follows that
\begin{align}
  \mathcal{F}^{(p+1)}\Big|=d A^{(p)}\Big|= \sum_{m=0}^{p} a_{p-m,m}dy^{m}_{a_{1\rightarrow p-m}\a_{1\rightarrow m}} (dx^a)^{1\rightarrow p-m}(d \q^{\a})^{1\rightarrow m}
\end{align}
because $b^{(p+1)}\big|=0$, which is readily deduced from \eqref{bresult}.

Finally, since $\mathcal{F}^{(p+1)}$ is, by construction, invariant under supersymmetry transformations, it must be constructed from the Maurer-Cartan forms and we ultimately conclude that
\begin{align}
 \mathcal{F}^{(p+1)}=\sum_{k=0}^{p} a_{p-k,k} E^{k}_{a_{1\rightarrow p-k}\a_{1\rightarrow k}}(E^a)^{1\rightarrow p-k}(E^\a)^{1\rightarrow k} \label{F}
\end{align}
where the constants $a_{p-k,k}$ are given by \eqref{ap0}-\eqref{app}.

Explicitly we find, for example, that the above result yields the following expressions:
\begin{align}
  \mathcal{F}^{(3)}&=\frac{2 \ii}{3}E^0_{ab}E^aE^b-\frac{3 \ii}{5}E^1_{a\a}E^aE^\a-\frac{2 \ii}{15}E^2_{\a\b}E^\a E^\b \\
  \mathcal{F}^{(4)}&=\frac{\ii}{2}E^0_{abc}E^aE^bE^c+\frac{87 \ii}{140}E^1_{ab\a}E^a E^b E^\a-\frac{9 \ii}{35}E^2_{a\a\b}E^a E^\a E^\b-\frac{3\ii}{35}E^3_{\a\b\g}E^\a E^\b E^\g \\
  \mathcal{F}^{(5)}&=\frac{2\ii}{5}E^0_{abcd}E^aE^bE^cE^d-\frac{13 \ii}{21}E^1_{abc\a}E^a E^b E^c E^\a-\frac{38 \ii}{105}E^2_{ab\a\b}E^aE^b E^\a E^\b \no \\ &\phantom{=}+\frac{8\ii}{35}E^3_{a\a\b\g}E^aE^\a E^\b E^\g+\frac{4\ii}{105}E^4_{\a\b\g\l}E^\a E^\b E^\g E^\l\\
   \mathcal{F}^{(6)}&=\frac{\ii}{3}E^0_{abcde}E^aE^bE^cE^dE^e+\frac{281 \ii}{462}E^1_{abcd\a}E^a E^b E^c E^d E^\a-\frac{104 \ii}{231}E^2_{abc\a\b}E^aE^b E^c E^\a E^\b \no \\ &\phantom{=}-\frac{94\ii}{231}E^3_{a b\a\b\g}E^a E^b E^\a E^\b E^\g+\frac{10\ii}{77}E^4_{a\a\b\g\l}E^a E^\a E^\b E^\g E^\l\no \\ &\phantom{=}+\frac{8\ii}{231}E^5_{\a\b\g\l\r}E^a E^\a E^\b E^\g E^\l E^\r\,.
\end{align}
Taking into account notational differences, $\mathcal{F}^{(3)}$ coincides with that given in \cite{Bergshoeff:1995hm}.  The general expression for $\mathcal{F}^{(p+1)}$ above reproduces $\mathcal{F}^{(3)}$ and $\mathcal{F}^{(4)}$ from our earlier work \cite{grassomcarthur},
and extends the known results to all valid $p>3$.

\section{Conclusions}\label{sec:conclusion}
\setcounter{equation}{0}

In this paper, we have used the integrability of the supersymmetry transformations assigned to $A^{(p)}$ in solving the cohomology problem
\be
h^{(p+2)} = d b^{(p+1)}, \quad \d_{\e} b^{(p+1)} = d \d_{\e} A^{(p)}\,,
\ee
to determine the superalgebras underlying $p$-brane actions for arbitrary values of $p$. Previously, these algebras have been derived on a case by case basis using a number of different techniques. The advantage of our approach is that is systematic and can be implemented for all values of $p$ in a single calculation. We have used the final compact expression for the superalgebras to obtain expressions for $\mathcal{F}^{(p+1)}=d A^{(p)}-b^{(p+1)}$, extending the known results to all valid $p$.

\begin{footnotesize}

\end{footnotesize}

\end{document}